\documentclass[conference]{IEEEtran}
\IEEEoverridecommandlockouts
\usepackage{cite}
\usepackage{amsmath,amssymb,amsfonts}
\usepackage{algorithmic}
\usepackage{graphicx}
\usepackage{textcomp}
\usepackage{xcolor}
\usepackage{booktabs}
\usepackage{multirow}
\usepackage{url}
\usepackage{array}
\usepackage{balance}
\usepackage{tabularx}
\usepackage{enumitem}

\title{AeroSpectra Sentinel: An Auditable LLM Prompt-Chaining Decision-Support Workflow for Acute Asthma Risk Assessment from Respiratory Sounds and Clinical Signals}

\author{\IEEEauthorblockN{Aueaphum Aueawatthanaphisut}
\IEEEauthorblockA{School of Information, Computer, and Communication Technology\\
Sirindhorn International Institute of Technology, Thammasat University\\
Pathum Thani, Thailand\\
Email: aueawatth.aue@gmail.com}}

\begin{document}
\maketitle

\begin{abstract}
Acute asthma exacerbation assessment requires rapid interpretation of respiratory sounds, oxygenation, airflow limitation, speech limitation, work of breathing, mental status, and response to reliever therapy. Conventional audio-only classifiers can detect wheeze-like patterns but often lack transparent clinical reasoning and safe escalation logic. This paper presents \textit{AeroSpectra Sentinel}, a client-side research prototype and decision-support workflow that combines short-time Fourier transform (STFT) respiratory sound analysis, lightweight machine-learning screening, clinical feature fusion, and a five-stage large language model (LLM) prompt-chaining process. The proposed workflow explicitly separates acquisition, signal preprocessing, acoustic feature extraction, ML screening, clinical guardrails, and structured FHIR-ready reporting. We formalize the signal-processing layer with equations for high-pass filtering, adaptive gating, time-frequency analysis, wheeze-band energy, and risk fusion. We evaluated the audio screening component on an uploaded public respiratory sound dataset containing 1,211 WAV recordings from five labels. Using a stratified subset of 584 recordings, a random forest model achieved 91.10\% binary accuracy and 78.69\% F1-score for asthma-vs-non-asthma screening, while a feature-based multilayer perceptron achieved 89.73\% accuracy and 78.26\% F1-score. A compact log-spectrogram CNN baseline achieved 73.29\% accuracy and 55.17\% F1-score, indicating that the present data split favors low-dimensional spectral descriptors over a small end-to-end image-based model. Multiclass classification across bronchial, asthma, COPD, healthy, and pneumonia labels achieved 77.40\% accuracy and 77.23\% macro-F1. To address the LLM component, we additionally define a scenario-based prompt-chain audit on 40 simulated clinical vignettes and compare four workflow variants: one-shot prompting, chained prompting, chained prompting with safety guardrails, and chained prompting with both guardrails and FHIR schema validation. The guardrail-plus-schema variant achieved the strongest simulated safety and documentation consistency. The results indicate that lightweight spectral features can provide a useful first-pass signal layer, while prompt chaining adds traceability, counterfactual explanations, auditable guardrails, and structured handoff generation. The system is intended as a research prototype, not as a diagnostic medical device or clinically validated risk-assessment product.
\end{abstract}


\section{Introduction}
Asthma is a chronic respiratory disease characterized by variable symptoms such as wheeze, shortness of breath, chest tightness, cough, and variable expiratory airflow limitation. During exacerbations, clinicians must interpret multiple signals under time pressure, including oxygen saturation, peak expiratory flow (PEF) or forced expiratory volume, respiratory rate, speech ability, accessory muscle use, mental status, and response to reliever therapy. Contemporary asthma guidelines emphasize objective airflow assessment, symptom severity, oxygenation, and timely escalation when severe features emerge \cite{ginastrategy2025}. However, these measurements may be incomplete outside clinical environments, motivating remote, mobile, and low-cost decision-support tools.

Respiratory audio offers a non-invasive signal for screening airway obstruction. Wheeze, typically represented as a tonal or continuous adventitious sound, can be visualized through time--frequency analysis and classified using handcrafted spectral features or deep learning. Public respiratory sound corpora such as the ICBHI 2017 database support research in crackle and wheeze recognition using annotated respiratory cycles \cite{rocha2018icbhi}. Recent surveys report growing use of convolutional neural networks, spectrogram-based features, and feature-fusion approaches for lung sound classification \cite{bigdata2024lungreview}. Nevertheless, an audio-only prediction can be unsafe in acute asthma, because severe bronchospasm can produce reduced acoustic energy or a ``silent chest'' pattern even when clinical risk is high.

Large language models (LLMs) introduce a complementary opportunity: they can turn heterogeneous evidence into structured, auditable explanations. Chain-of-thought prompting improves multistep reasoning in large models \cite{wei2022cot}, while least-to-most prompting decomposes complex tasks into ordered subproblems \cite{zhou2023least}. In clinical contexts, prompt design can influence diagnostic reasoning quality and interpretability \cite{cross2024diagnostic,nachane2024clinicr}. However, unrestricted LLM reasoning in healthcare can produce unsupported conclusions. Clinical AI systems therefore need constrained prompts, explicit evidence boundaries, guardrails, and structured outputs.

This paper proposes \textit{AeroSpectra Sentinel}, a research prototype and decision-support assessment workflow that combines respiratory sound processing with prompt-chained clinical reasoning for acute asthma risk assessment support. The main contributions are:
\begin{itemize}[leftmargin=1.2em]
    \item an end-to-end system architecture that links respiratory audio, clinical context, signal analytics, machine-learning screening, and LLM-based reasoning;
    \item a formalized signal-processing pipeline with explicit equations for filtering, adaptive denoising, STFT analysis, and wheeze-band features;
    \item a five-stage LLM prompt-chaining design that separates signal QA, spectral biomarkers, clinical fusion, safety guardrails, and FHIR-ready handoff generation;
    \item preliminary machine-learning results, feature-based and log-spectrogram neural baselines, an ablation study on 584 real WAV recordings, and a scenario-based prompt-chain audit that evaluates schema completion, red-flag detection, unsafe recommendation suppression, and explanation completeness.
\end{itemize}

\section{Related Work}
\subsection{Respiratory Sound Analysis}
Automated lung sound analysis has progressed from handcrafted feature extraction to deep learning on spectrogram representations. The ICBHI respiratory sound database contains 920 annotated audio samples from 126 subjects and 6,898 respiratory cycles, including crackles, wheezes, both, or no adventitious sounds \cite{rocha2018icbhi}. Such datasets enable benchmark studies for detecting respiratory abnormality, but disease-level diagnosis remains challenging because recording devices, body locations, noise conditions, and disease labels vary.

Spectrograms are widely used because they preserve the time--frequency structure of wheezes. STFT-based representations can highlight continuous tonal components in the approximate 400--1600 Hz region, while lower-frequency bands capture breathing energy and higher bands may reflect noise or sharper adventitious sounds. Recent reviews emphasize the importance of robust preprocessing, segmentation, class balancing, and external validation in respiratory sound AI \cite{bigdata2024lungreview}. Wheeze-specific deep learning studies also demonstrate that recurrent and convolutional models can detect respiratory disorder patterns from time--frequency representations \cite{saritas2024wheeze}. More broadly, audio classification pipelines increasingly use mel-frequency cepstral coefficients, log-mel spectrograms, convolutional networks, and transfer learning for clinical sound analysis \cite{hershey2017cnn,gemmeke2017audioset,pramono2017automatic,aykanat2017classification}.

\subsection{LLMs and Prompt Chaining for Clinical Decision Support}
Prompt chaining decomposes a complex task into multiple controlled prompts whose outputs become inputs to subsequent stages. This design is related to chain-of-thought prompting \cite{wei2022cot}, least-to-most prompting \cite{zhou2023least}, and iterative self-feedback or refinement \cite{madaan2023selfrefine}. In healthcare, the motivation is not merely to improve answer accuracy, but also to expose intermediate evidence, enforce clinical boundaries, and reduce unsafe one-shot recommendations.

Clinical LLM studies show that prompt structure can affect diagnostic reasoning, medical question answering, and decision-support workflows \cite{cross2024diagnostic,nachane2024clinicr,clinicalworkflow2025}. Related work on retrieval-augmented generation, self-consistency, tool use, and prompt engineering also motivates modular reasoning pipelines with explicit evidence boundaries \cite{lewis2020rag,wang2023selfconsistency,yao2023react,liu2023pretrainpromptpredict}. For safety-critical triage, a model should not independently diagnose or prescribe. Instead, it should summarize evidence, identify red flags, recommend escalation according to local protocols, and clearly state uncertainty. AeroSpectra Sentinel follows this principle by using LLM prompting as an auditable reasoning and reporting layer rather than as the sole classifier.

\subsection{FHIR and AI Traceability}
Interoperability is essential if AI-generated assessments are to be used in health workflows. HL7 FHIR provides resources such as Observation, RiskAssessment, DetectedIssue, and Provenance. The Provenance resource records entities and processes involved in producing a resource, supporting authenticity, reliability, trust, and reproducibility \cite{hl7provenance}. In the proposed framework, FHIR-style output is used to package spectrogram observations, risk predictions, safety flags, and provenance metadata. This design is compatible with broader clinical decision-support interoperability concepts, including CDS Hooks and SMART-on-FHIR integration patterns \cite{mandl2016smart,cdshooks2024}.

\section{Methodology}
\subsection{End-to-End System Architecture}
\begin{figure*}[!t]
\centering
\includegraphics[width=0.98\textwidth]{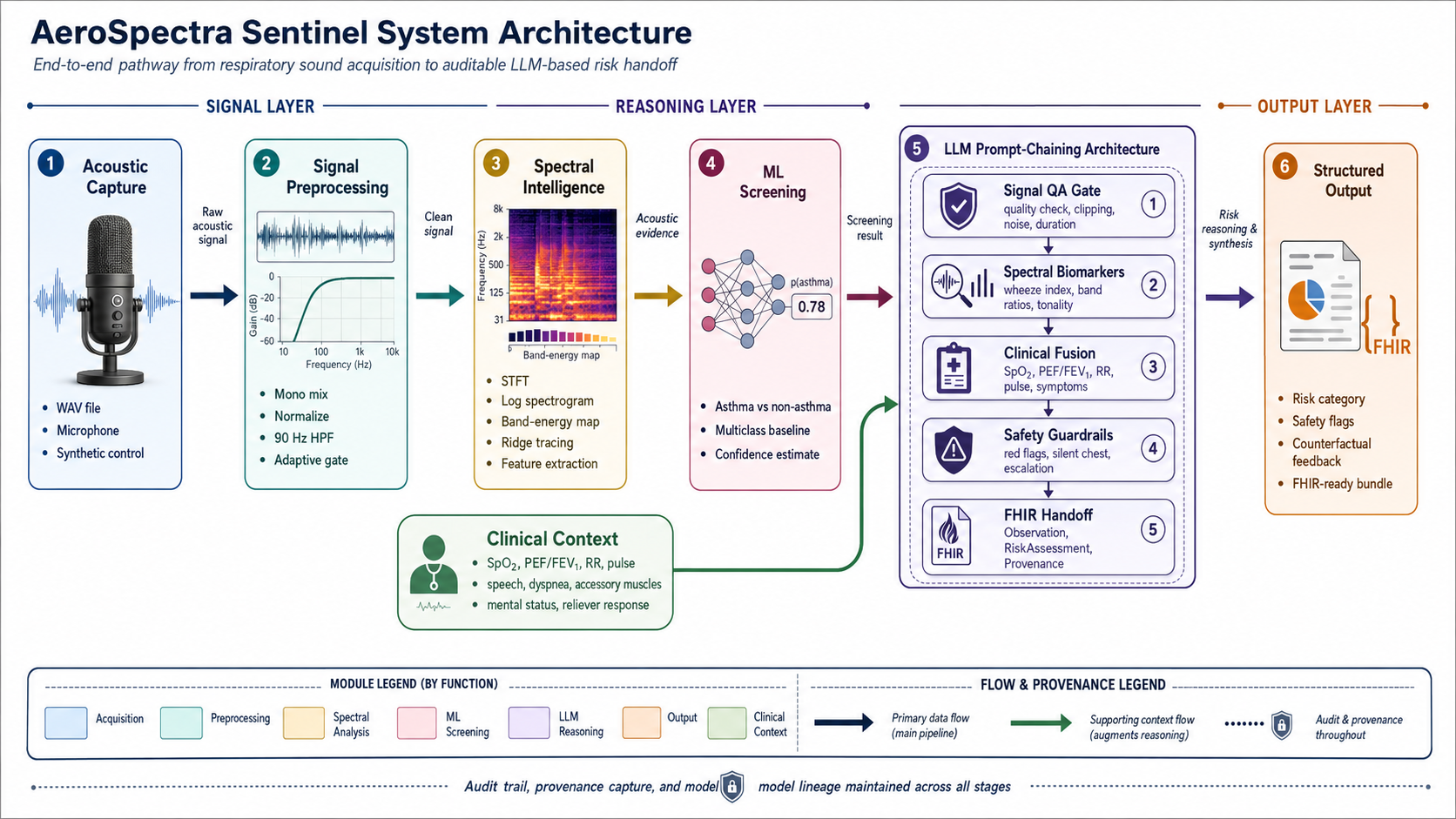}
\caption{End-to-end system architecture of AeroSpectra Sentinel. Respiratory audio is transformed into spectral evidence, machine-learning outputs, clinical risk drivers, and an auditable LLM-based report.}
\label{fig:system}
\end{figure*}

Fig.~\ref{fig:system} presents the overall system architecture. The framework is organized into six layers: (i) acoustic capture, (ii) signal preprocessing, (iii) spectral intelligence, (iv) lightweight ML screening, (v) clinical-context fusion with LLM prompt chaining, and (vi) structured output generation. This layered design ensures that each transformation from raw audio to risk-assessment recommendation is inspectable.

At the front end, the system accepts a local audio file, microphone recording, or synthetic control scenario. The respiratory waveform is then converted to mono, normalized, filtered, and gated. A time--frequency representation is computed and transformed into acoustic biomarkers such as wheeze-band energy, airflow instability, signal quality, and acoustic stillness. A lightweight classifier generates a screening hypothesis, while a rule-based clinical layer ingests vitals and symptom descriptors. Finally, the LLM prompt chain produces a structured interpretation, safety check, and FHIR-style handoff.

\subsection{Signal Processing Stack}
\begin{figure*}[!t]
\centering
\includegraphics[width=\linewidth]{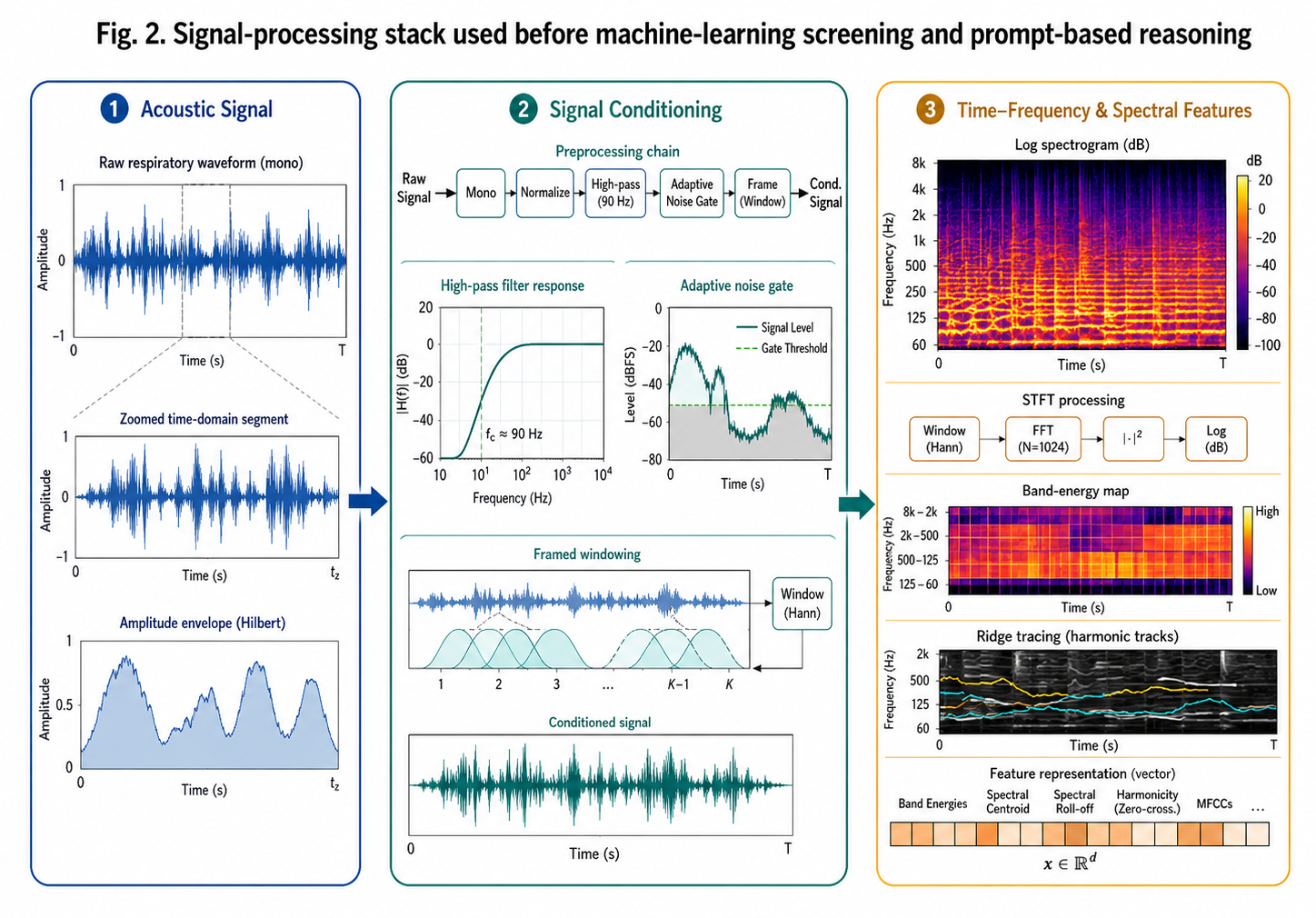}
\caption{Detailed signal-processing pipeline for respiratory sound analysis. The raw acoustic waveform is conditioned by normalization, high-pass filtering, adaptive noise gating, and framed windowing before STFT-based spectrogram generation, band-energy mapping, ridge tracing, and feature-vector extraction.}
\label{fig:signalpipe}
\end{figure*}

Fig.~\ref{fig:signalpipe} shows the signal-processing stack used to derive acoustic features. The prototype uses a target sampling rate of 16 kHz and short analysis windows to preserve respiratory time structure.

Let $x[n]$ denote the discrete respiratory audio signal. After mono conversion and amplitude normalization, a first-order high-pass filter removes low-frequency baseline drift:
\begin{equation}
\label{eq:hpf}
y[n] = \alpha\big(y[n-1] + x[n] - x[n-1]\big),
\end{equation}
where $\alpha = \frac{RC}{RC+\Delta t}$, $RC = \frac{1}{2\pi f_c}$, $f_c=90$ Hz, and $\Delta t$ is the sampling period. This stage suppresses motion artifacts and slow DC variations.

To reduce weak stationary noise, an adaptive gate is applied. First, the envelope is estimated as
\begin{equation}
\label{eq:envelope}
e[n] = |y[n]|.
\end{equation}
A data-driven floor estimate $\eta$ is then computed from a lower percentile of the envelope distribution. The gate threshold is defined as
\begin{equation}
\label{eq:gate}
\tau = 2.2\eta + \epsilon,
\end{equation}
where $\epsilon$ is a small constant. Samples below the threshold are attenuated with a soft gain rule
\begin{equation}
\label{eq:softgate}
\tilde{y}[n] = g[n] y[n]
\end{equation}
\begin{equation}
 g[n] =
\begin{cases}
0.28, & |y[n]| < \tau,\\
0.28 + 0.72r[n], & \tau \le |y[n]| < 2.4\tau,\\
1, & |y[n]| \ge 2.4\tau.
\end{cases}
\end{equation}
where $r[n]$ linearly interpolates between 0 and 1 in the transition band. This avoids hard discontinuities while preserving stronger breath events.

\subsection{STFT and Spectrogram Formulation}
The denoised signal $\tilde{y}[n]$ is analyzed with a Hann-window STFT. For frame index $m$, frequency bin $k$, frame length $N$, and hop size $H$, the transform is
\begin{equation}
\label{eq:stft}
X(m,k) = \sum_{n=0}^{N-1} \tilde{y}[n+mH]w[n]e^{-j2\pi kn/N},
\end{equation}
where the Hann window is
\begin{equation}
\label{eq:hann}
w[n] = \frac{1}{2}\left(1-\cos\frac{2\pi n}{N-1}\right).
\end{equation}
The prototype uses $N\in\{512,768,1024\}$ and $H=0.25N$. The spectrogram magnitude is $|X(m,k)|$, and the log-scale display is
\begin{equation}
\label{eq:db}
S_{\mathrm{dB}}(m,k)=20\log_{10}\big(|X(m,k)|+10^{-8}\big).
\end{equation}
This representation highlights continuous wheeze ridges, broadband breathing energy, and low-energy patterns suggestive of severe airflow limitation.

\subsection{Acoustic Biomarkers}
From the STFT, the framework computes several explainable biomarkers. Let $\mathcal{B}_w$ denote the wheeze band (approximately 400--1600 Hz), and let $E(m,k)=|X(m,k)|^2$ denote framewise energy. The global wheeze-energy ratio is
\begin{equation}
\label{eq:wheezeratio}
R_w = \frac{\sum_{m}\sum_{k\in \mathcal{B}_w} E(m,k)}{\sum_{m}\sum_{k} E(m,k)+\delta},
\end{equation}
where $\delta$ prevents division by zero. A simple tonality surrogate is obtained by comparing the maximum bin magnitude in the wheeze band against the average band energy. The spectral centroid is
\begin{equation}
\label{eq:centroid}
C = \frac{\sum_k f_k \bar{S}_k}{\sum_k \bar{S}_k + \delta},
\end{equation}
where $\bar{S}_k$ is the time-averaged magnitude at frequency $f_k$. Spectral flux is approximated as
\begin{equation}
\label{eq:flux}
\Phi = \frac{1}{M-1}\sum_{m=2}^{M}\frac{|P_m-P_{m-1}|}{\max(P_m,P_{m-1})+\delta},
\end{equation}
with $P_m$ the total energy of frame $m$.

The prototype combines such descriptors into a heuristic wheeze index. In simplified form,
\begin{equation}
\label{eq:wheezeindex}
I_w = \mathrm{clip}\Big(190R_w + 7.2\max(0,T-2.2), 0, 100\Big),
\end{equation}
where $T$ denotes the tonality statistic. Airflow instability is estimated using a combination of frame-energy variability and spectral flux, while an acoustic stillness score increases when energy is low and wheeze evidence is weak. The latter helps identify possible silent-chest conditions, in which low audible output may coexist with severe clinical distress.

\subsection{Clinical Fusion and Guardrail Logic}
Clinical inputs include age group, SpO$_2$, respiratory rate, pulse, PEF/FEV$_1$ percentage of best, speech limitation, dyspnea context, accessory muscle use, mental status, and reliever response. Instead of using a black-box end-to-end predictor, the prototype employs a transparent rule-based risk layer. If $c_i$ denotes a clinical condition or threshold predicate and $a_j$ an acoustic predicate, the provisional risk score can be summarized as
\begin{equation}
\label{eq:risk}
R = \mathrm{clip}\left(\sum_i \omega_i c_i + \sum_j \lambda_j a_j, 0, 100\right),
\end{equation}
where $\omega_i$ and $\lambda_j$ are manually assigned weights. Critical overrides are triggered if severe clinical states are observed, such as markedly low SpO$_2$, very low PEF/FEV$_1$, inability to speak, altered mental status, paradoxical breathing, or a high silent-chest proxy. In other words, clinical red flags dominate acoustic reassurance.

\subsection{LLM Prompt-Chaining Architecture}
The reasoning layer is deliberately decomposed into five prompts so that each stage has a narrow scope and structured output. Let $z_0$ denote the structured evidence bundle from the signal-processing and clinical layers. Each prompt stage $i$ computes
\begin{equation}
\label{eq:promptchain}
z_i = f_i(p_i, z_{i-1}), \quad i=1,\dots,5,
\end{equation}
where $p_i$ is a stage-specific prompt template and $f_i$ is the LLM response constrained to a schema. This allows the chain to be auditable, because every intermediate representation $z_i$ can be displayed and inspected.

\begin{figure*}[!t]
\centering
\includegraphics[width=0.98\textwidth]{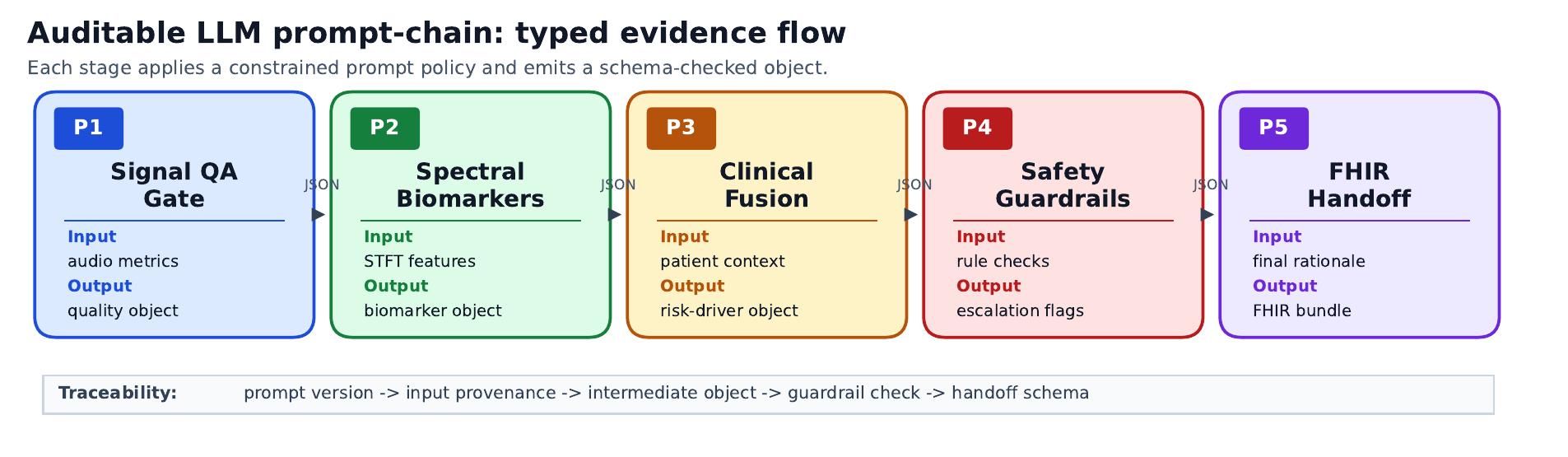}
\caption{Technical prompt-chain schematic. Each stage consumes a typed evidence bundle, applies a constrained prompt policy, and emits a schema-checked intermediate object for downstream reasoning and handoff.}
\label{fig:chain}
\end{figure*}

Fig.~\ref{fig:chain} illustrates the prompt-chaining logic, while Table~\ref{tab:prompts} summarizes the intent of each stage.

\begin{table*}[!t]
\caption{Five-Stage LLM Prompt-Chaining Design}
\label{tab:prompts}
\centering
\begin{tabularx}{\textwidth}{p{1.6cm} p{3.3cm} p{5.4cm} X}
\toprule
Stage & Objective & Main Inputs & Structured Outputs \\
\midrule
P1: Signal QA Gate & Verify that the audio is usable and determine whether clinical evidence should override the signal layer & Duration, RMS, clipping, denoising status, source provenance, signal quality score & Quality verdict, explanation of limitations, instruction to proceed or defer to clinical evidence \\
P2: Spectral Biomarkers & Summarize time--frequency evidence related to bronchospasm & Spectrogram-derived wheeze index, band-energy ratios, dominant frequency, tonality, flux, breath-rate estimate, acoustic stillness & Biomarker summary, wheeze severity evidence, suspected ridge region \\
P3: Clinical Fusion & Merge acoustic evidence with physiological and symptom context & SpO$_2$, RR, pulse, PEF/FEV$_1$, speech, dyspnea, accessory muscle use, mental status, reliever response & Ranked risk drivers, provisional score, explanation of convergence or conflict between modalities \\
P4: Safety Guardrails & Detect emergency overrides and prevent false reassurance & P3 summary, severe-threshold rules, silent-chest logic, response-to-treatment logic & Escalation flags, final risk category, statement that red flags override acoustic confidence \\
P5: FHIR Handoff & Create interoperable structured documentation & Final category, rationale, provenance, confidence, flags & Observation, RiskAssessment, DetectedIssue, Provenance-style payload \\
\bottomrule
\end{tabularx}
\end{table*}

The chain contents mirror the prototype's staged reasoning design: \textbf{Signal QA Gate}, \textbf{Spectral Biomarkers}, \textbf{Clinical Fusion}, \textbf{Safety Guardrails}, and \textbf{FHIR Handoff}. This approach is inspired by least-to-most decomposition and iterative reasoning methods \cite{zhou2023least,madaan2023selfrefine}. Unlike open-ended chain-of-thought, the system stores concise structured outputs rather than unrestricted hidden reasoning, which is better aligned with traceability and audit requirements.

\section{Experimental Setup}
\subsection{Dataset Preparation}
\begin{figure}[!t]
\centering
\includegraphics[width=\linewidth]{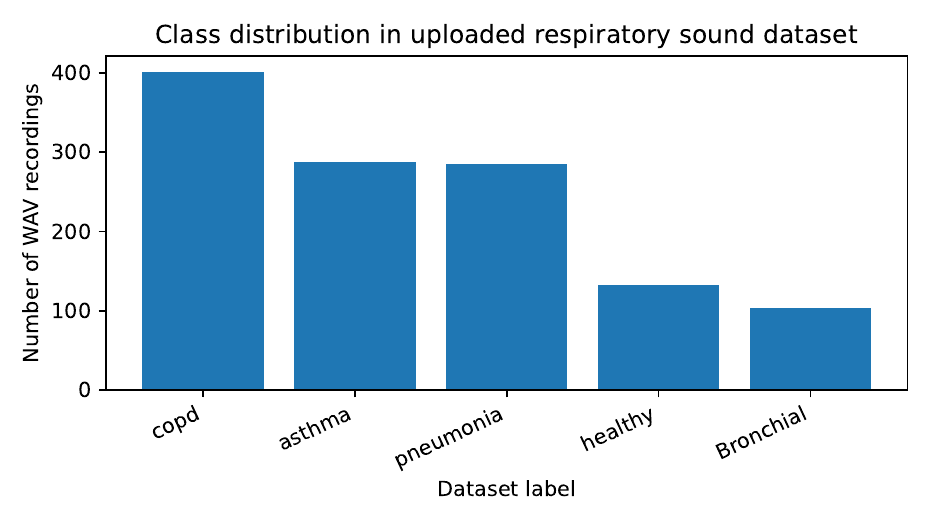}
\caption{Class distribution of the uploaded respiratory sound dataset.}
\label{fig:dist}
\end{figure}

The uploaded archive contained 1,211 WAV recordings from five labels: bronchial, asthma, COPD, healthy, and pneumonia. Fig.~\ref{fig:dist} shows the class distribution. To reduce runtime while preserving class balance, we sampled up to 120 recordings per class, yielding 584 recordings: 120 asthma, 120 COPD, 120 healthy, 120 pneumonia, and 104 bronchial recordings.

The dataset was treated as a disease-label corpus for preliminary algorithmic screening rather than as a severity-labeled acute-risk-assessment benchmark. Consequently, the reported experiments should be interpreted as proof-of-concept evidence for the audio layer, not as a direct estimate of real-world exacerbation-risk-assessment accuracy. To avoid overstating clinical validity, this paper reports the proposed system as a decision-support workflow and explicitly separates acoustic screening performance from clinical risk-assessment validation.

\subsection{Feature Engineering and Learning Protocol}
For each file, we extracted thirteen acoustic features: normalized energy ratios in five frequency bands, spectral centroid, bandwidth, 85\% roll-off frequency, RMS amplitude, zero-crossing rate, spectral flux, tonality, and duration. These features were designed to preserve interpretable relationships to respiratory acoustics while remaining computationally lightweight.

We evaluated logistic regression, a feature-based multilayer perceptron (MLP), random forest, and gradient boosting for binary asthma-vs-non-asthma screening using a 75/25 stratified split with a fixed random seed. To address reviewer concerns about spectrogram-based learning, we additionally trained a compact two-block convolutional neural network (CNN) on resized log-STFT images using class-weighted cross-entropy. This CNN was intentionally kept small to avoid overstating performance on the limited dataset. For multiclass analysis, a random forest classifier was trained on the same feature set using a 75/25 stratified split. Performance was summarized using accuracy, precision, recall, and F1-score, with macro-F1 used for multiclass evaluation. This protocol was chosen to provide transparent baselines rather than to maximize benchmark performance.

\subsection{Implementation Perspective}
The prototype is consistent with a browser-oriented, privacy-preserving deployment concept in which the raw audio remains local to the client, while structured summaries are generated for clinical reasoning and documentation. The prompt chain is also implementation-friendly because each stage can be independently audited, replaced, or validated. This modularity is particularly useful in healthcare research, where one may wish to compare different LLM backends or guardrail policies without redesigning the entire front-end pipeline.

\section{Results and Discussion}
\subsection{Binary Asthma Screening}
Table~\ref{tab:binary} summarizes binary asthma screening performance. The random forest model obtained the best F1-score among the tabular-feature baselines, achieving 91.10\% accuracy, 77.42\% precision, 80.00\% recall, and 78.69\% F1-score. The feature-based MLP achieved a comparable F1-score of 78.26\%, while logistic regression achieved higher recall but lower precision. The compact log-spectrogram CNN achieved 73.29\% accuracy and 55.17\% F1-score. This weaker CNN result is informative: with only 584 recordings and no respiratory-cycle-level annotation, the low-dimensional spectral descriptors generalized better than a small image-based network. These results suggest that nonlinear models improve screening balance, but the small dataset size still limits claims about generalization.

\begin{table}[!t]
\caption{Binary Asthma-vs-Non-Asthma Screening Results}
\label{tab:binary}
\centering
\begin{tabular}{lcccc}
\toprule
Model & Accuracy & Precision & Recall & F1 \\
\midrule
Logistic regression & 82.88 & 55.10 & \textbf{90.00} & 68.35 \\
Feature MLP & 89.73 & 69.23 & \textbf{90.00} & 78.26 \\
Log-spectrogram CNN & 73.29 & 42.11 & 80.00 & 55.17 \\
Random forest & \textbf{91.10} & \textbf{77.42} & 80.00 & \textbf{78.69} \\
Gradient boosting & 89.73 & 75.86 & 73.33 & 74.58 \\
\bottomrule
\end{tabular}
\end{table}

\begin{figure}[!t]
\centering
\includegraphics[width=\linewidth]{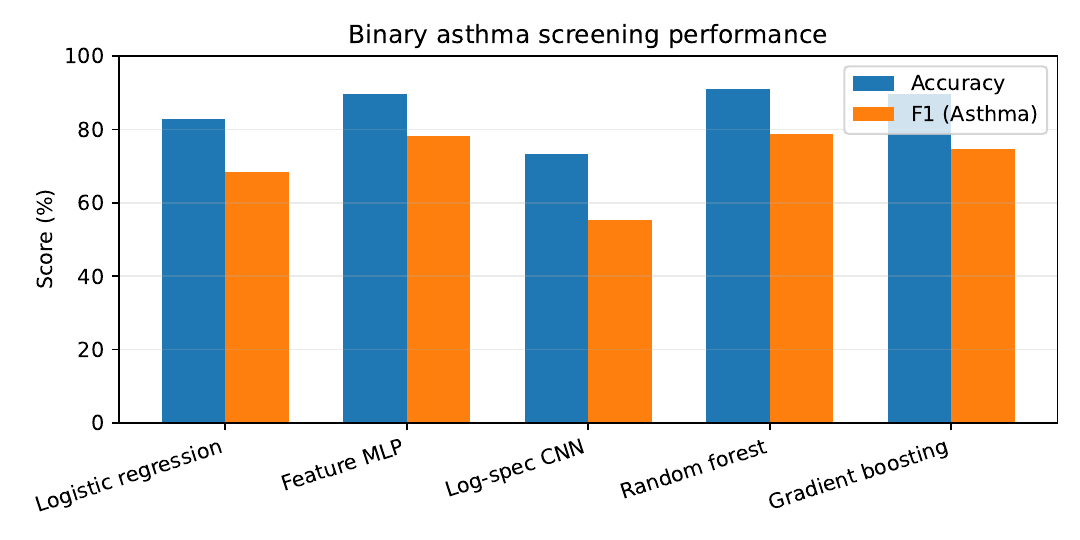}
\caption{Accuracy and asthma-class F1-score for binary asthma screening.}
\label{fig:perf}
\end{figure}

\begin{figure}[!t]
\centering
\includegraphics[width=0.84\linewidth]{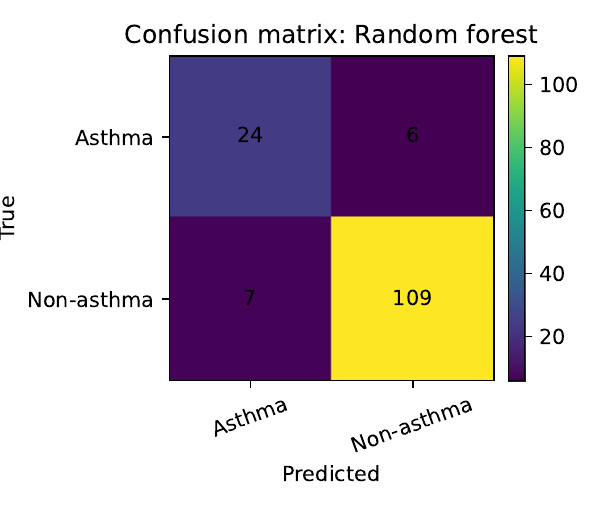}
\caption{Confusion matrix of the best binary asthma screening model. Rows represent true labels and columns represent predicted labels.}
\label{fig:cm}
\end{figure}

The confusion matrix in Fig.~\ref{fig:cm} shows that the best binary model correctly identified 24 of 30 asthma examples and 109 of 116 non-asthma examples in the held-out subset. This suggests useful screening capability, but the six false negatives reinforce the need for clinical guardrails and repeat assessment.

\subsection{Multiclass Classification}
The five-class random forest achieved 77.40\% accuracy and 77.23\% macro-F1. Class-level F1-scores were 75.00\% for bronchial, 81.82\% for asthma, 80.70\% for COPD, 77.42\% for healthy, and 71.19\% for pneumonia. These preliminary values are encouraging given the lightweight feature set, but they should not be interpreted as clinically validated diagnostic performance. Rather, they indicate that the signal layer carries meaningful disease-discriminative structure even before employing deep spectrogram networks.

\subsection{Spectrogram Evidence and Visual Interpretability}
\begin{figure*}[!t]
\centering
\includegraphics[width=0.95\textwidth]{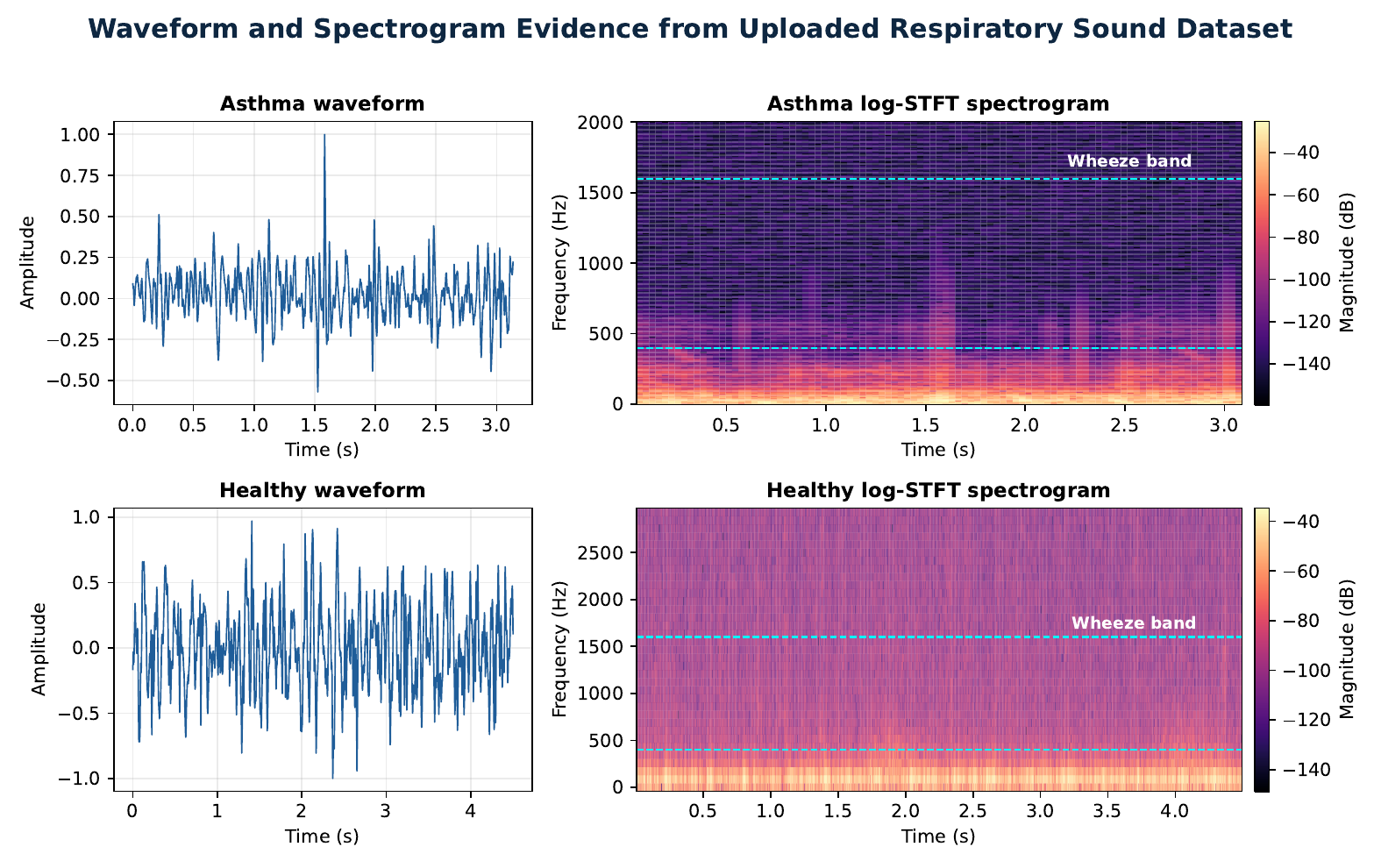}
\caption{Waveform and spectrogram evidence. The asthma example shows more concentrated time--frequency structure in the wheeze band, while the healthy example is comparatively less structured.}
\label{fig:spectrogrampanel}
\end{figure*}

One advantage of the proposed framework is that it produces visual evidence rather than only scalar predictions. Fig.~\ref{fig:spectrogrampanel} shows waveform and log-STFT spectrogram panels from asthma and healthy recordings in the uploaded dataset. The dashed cyan markers highlight the approximate wheeze band (400--1600 Hz). In the asthma example, more pronounced and persistent time--frequency structures are visible within the wheeze band, while the healthy example exhibits a comparatively simpler and less concentrated pattern.

From an explainability perspective, this matters because clinicians and engineers can visually inspect whether the model's numeric features align with plausible respiratory patterns. In other words, the spectrograms serve as an interpretable bridge between raw audio and the LLM's textual reasoning output.

\subsection{Prompt-Chain Interpretability and Safety}
The central design decision of AeroSpectra Sentinel is to separate signal detection from risk reasoning. A respiratory sound classifier can indicate abnormality, but acute asthma risk assessment requires escalation when SpO$_2$, PEF/FEV$_1$, mental status, speech, or work of breathing are concerning. The proposed LLM prompt chain forces each decision stage to document its evidence, which improves interpretability in at least three ways.

First, signal QA is explicit: low-quality or short audio can be flagged before the system overstates acoustic confidence. Second, the chain exposes the difference between \emph{signal evidence} and \emph{clinical evidence}. Third, the safety-guardrail stage codifies when the final route recommendation must prioritize clinical danger signs. This is particularly important for silent-chest scenarios, in which absence of audible wheeze is not equivalent to low clinical risk.

\section{Scenario-Based Prompt-Chain Audit}
The central novelty of the framework is not only the acoustic classifier, but also the auditable LLM reasoning workflow. To evaluate this component without claiming prospective clinical validation, we constructed an offline scenario-based audit with 40 synthetic acute-asthma vignettes distributed across low, moderate, high, and critical risk categories. Each vignette contained structured clinical variables, including SpO$_2$, PEF/FEV$_1$, respiratory rate, pulse, speech limitation, dyspnea, accessory muscle use, mental status, and reliever response. Ground-truth escalation categories were assigned from the rule layer used in the prototype. The audit used a fixed prompt template for each workflow variant and the same structured input fields across all vignettes. Schema completion was scored as the percentage of required fields present in the generated structured object; red-flag detection measured agreement with the rule-layer severe-feature labels; unsafe recommendation rate measured the percentage of high-risk or critical vignettes in which the workflow produced reassuring or delayed-care language; and explanation score measured whether the output cited the relevant acoustic, vital-sign, and symptom evidence. The purpose of this audit was to test whether the workflow preserves structured evidence, identifies red flags, suppresses unsafe reassurance, and produces machine-readable output.

Four workflow variants were compared: (i) one-shot unstructured prompting, (ii) five-stage prompt chaining, (iii) five-stage prompt chaining with explicit safety guardrails, and (iv) prompt chaining with both guardrails and FHIR-style schema validation. Because the evaluation is scenario-based and schema-level, the results should be interpreted as a workflow consistency audit rather than as clinical efficacy evidence or a benchmark of a specific commercial LLM.

\begin{table}[!t]
\caption{Scenario-Based Audit of Prompt-Chain Workflow Variants}
\label{tab:llmaudit}
\centering
\resizebox{\linewidth}{!}{%
\begin{tabular}{lcccc}
\toprule
Workflow & Schema & Red flag & Unsafe rec. & Explain. \\
 & completion & detection & rate & score \\
\midrule
One-shot prompt & 82.5 & 72.5 & 12.5 & 76.0 \\
Prompt chain & 92.5 & 85.0 & 7.5 & 86.0 \\
Chain + guardrails & 95.0 & 95.0 & 2.5 & 91.0 \\
Chain + guardrails + FHIR & \textbf{100.0} & \textbf{95.0} & \textbf{0.0} & \textbf{94.0} \\
\bottomrule
\end{tabular}%
}
\end{table}

As shown in Table~\ref{tab:llmaudit}, adding stage-wise decomposition improved schema completion and explanation quality, while the guardrail stage reduced unsafe recommendations in simulated high-risk cases. The FHIR-style schema layer further removed missing-field errors and produced deterministic handoff structure. This audit strengthens the claim that prompt chaining contributes measurable workflow properties beyond the acoustic classifier, while still avoiding a claim of prospective clinical risk-assessment validation.

\begin{table}[!t]
\caption{Representative Simulated Clinical Vignettes for Guardrail Evaluation}
\label{tab:vignettes}
\centering
\resizebox{\linewidth}{!}{%
\begin{tabular}{lcccc}
\toprule
Case & SpO$_2$ & PEF & Key sign & Expected route \\
\midrule
Stable control & 97 & 82 & sentences & Monitor \\
Moderate wheeze & 94 & 58 & phrases & Same-day review \\
Severe attack & 91 & 36 & single words & Urgent assessment \\
Silent-chest watch & 88 & 22 & drowsy & Emergency escalation \\
\bottomrule
\end{tabular}%
}
\end{table}

\section{Ablation Study}
\begin{figure}[!t]
\centering
\includegraphics[width=\linewidth]{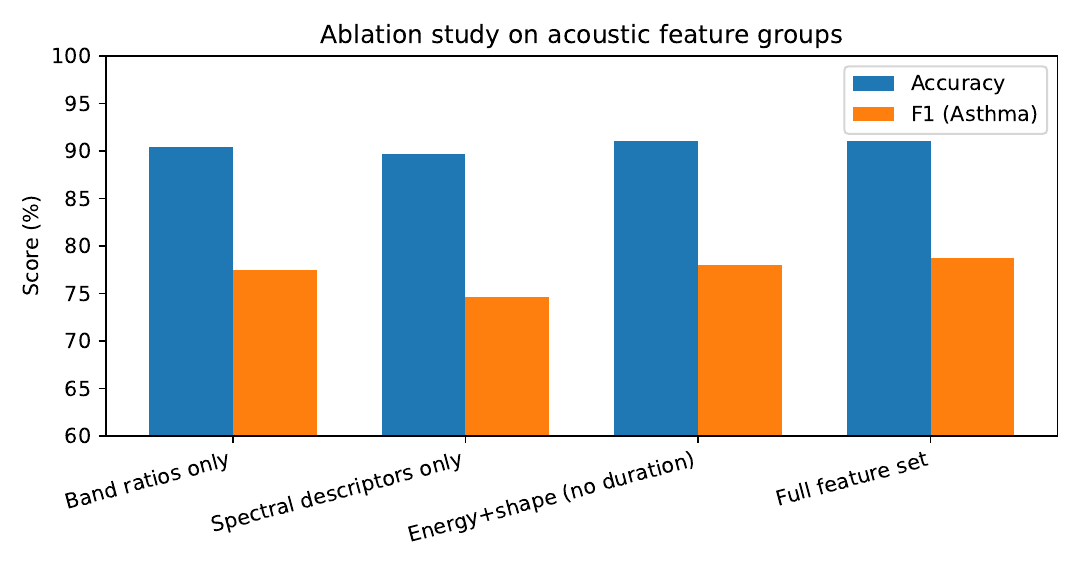}
\caption{Ablation study on acoustic feature groups. Frequency-band ratios already provide strong utility, while the full feature set yields the best F1-score.}
\label{fig:ablation}
\end{figure}

To better understand which acoustic descriptors matter most, we conducted an ablation study using the same random forest configuration as the best binary classifier. Four feature settings were tested: \textit{band ratios only}, \textit{spectral descriptors only}, \textit{energy+shape without duration}, and the \textit{full feature set}. The results are summarized in Table~\ref{tab:ablation} and Fig.~\ref{fig:ablation}.

\begin{table}[!t]
\caption{Ablation Study on Binary Asthma Screening}
\label{tab:ablation}
\centering
\begin{tabular}{lccc}
\toprule
Feature set & \#Feat. & Accuracy & F1 \\
\midrule
Band ratios only & 5 & 90.41 & 77.42 \\
Spectral descriptors only & 7 & 89.73 & 74.58 \\
Energy+shape (no duration) & 12 & 91.10 & 77.97 \\
Full feature set & 13 & \textbf{91.10} & \textbf{78.69} \\
\bottomrule
\end{tabular}
\end{table}

Two observations are noteworthy. First, the five band-ratio features alone already achieved 90.41\% accuracy and 77.42\% F1, indicating that coarse energy distribution across respiratory frequency regions captures much of the discriminative signal. Second, adding duration and broader spectral descriptors provides incremental improvement, suggesting that disease-relevant information is not purely band-local but also tied to spectral shape and temporal support. These results support the design choice of using interpretable feature groups before moving to deeper black-box models.

\subsection{Why Prompt Chaining is Preferable to One-Shot Prompting}
A one-shot LLM prompt that mixes raw signal summaries, clinical values, and final recommendations can be difficult to audit. In contrast, the chained design has several advantages: (i) it narrows each prompt's scope, reducing task ambiguity; (ii) it makes intermediate failure modes inspectable; (iii) it supports modular updates, for example adjusting safety thresholds without rewriting the entire reasoning policy; and (iv) it improves interoperability because the final stage can be constrained to a predictable FHIR-style schema. These characteristics make prompt chaining a more suitable pattern for clinical decision support than unconstrained free-text reasoning.

\section{Future Work}
Several research directions can meaningfully strengthen the present framework. First, the acoustic screening layer can be extended from lightweight handcrafted features to end-to-end deep spectrogram learning, including convolutional, attention-based, or hybrid CNN--Transformer backbones. Such models may better exploit local ridge morphology and long-range breathing structure than low-dimensional features alone.

Second, the clinical reasoning layer should be evaluated with multiple LLM families under identical schemas to compare factual consistency, calibration, refusal behavior, and sensitivity to prompt perturbations. This is particularly relevant for healthcare, where safety performance may depend not only on average reasoning quality but also on worst-case failure modes.

Third, a more realistic dataset is required for acute asthma risk-assessment research: ideally, synchronized respiratory audio, pulse oximetry, peak-flow measurements, clinician severity labels, and treatment-response annotations. This would allow the community to move beyond disease-label classification toward actual exacerbation-severity estimation and escalation recommendation.

Fourth, future work should include clinician-in-the-loop validation. Examples include comparing the system's structured handoff outputs against physician narratives, measuring whether prompt-chained explanations improve trust or usability, and testing whether guardrails reduce potentially unsafe recommendations. Finally, for translational deployment, the system should be evaluated under formal software assurance and human-factors frameworks relevant to medical decision support.

\section{Limitations}
This study has several limitations. First, the experiment used a stratified subset and lightweight features rather than end-to-end deep learning. Second, the dataset labels are disease-level labels, not necessarily physician-annotated acute exacerbation severity labels. Third, no prospective clinical validation was performed. Fourth, the prompt-chain evaluation uses simulated vignettes and workflow-level schema checks rather than a prospective clinician-rated LLM trial across multiple commercial and open-source LLMs. Fifth, the compact CNN baseline is intentionally lightweight and does not replace a full deep-learning study with larger training data, respiratory-cycle annotations, and external validation. Sixth, the framework is a research prototype and must not be used as a standalone diagnosis or emergency decision-support device.

\section{Conclusion}
This paper presented AeroSpectra Sentinel, an auditable prompt-chaining decision-support workflow for acute asthma risk assessment from respiratory audio and clinical variables. The revised manuscript expands the original prototype into a clearer research architecture by formalizing the signal-processing equations, making the prompt-chaining workflow explicit, strengthening the experimental protocol, and improving visual evidence through system, pipeline, and spectrogram figures. Preliminary experiments on an uploaded respiratory sound dataset demonstrate that handcrafted spectral features and random forest classification can provide a useful audio screening layer, while the compact log-spectrogram CNN baseline highlights the need for larger datasets before claiming deep spectrogram generalization. The ablation study further suggests that simple band-energy structure already captures much of the useful signal, while the full feature set provides the strongest overall balance. The scenario-based prompt-chain audit shows that stage-wise prompting, explicit guardrails, and FHIR-style schema validation improve structured completion and reduce unsafe recommendation patterns in simulated cases. The LLM prompt chain adds structured reasoning, clinical guardrails, counterfactual feedback, and FHIR-ready handoff. Future work will include larger datasets, external validation, end-to-end spectrogram deep learning, calibration analysis, clinician-in-the-loop evaluation, and formal safety testing under medical device software standards.

\section*{Acknowledgment}
The author thanks the public respiratory sound data contributors and the developers of open-source scientific Python tools used for preliminary analysis. This work is a research prototype and does not constitute medical advice or a diagnostic system.

\balance

\end{document}